\documentclass[final,3p,times,11pt,onecolumn]{elsarticle}

\usepackage{pifont,bbm,amsfonts,mathrsfs,amsmath,stmaryrd,amssymb}
\usepackage{amscd,graphicx,array,dsfont,texdraw,tikz,footnote,verbatim}%
\usepackage[normalem]{ulem}

\makesavenoteenv{tabular}
\usetikzlibrary{arrows,shapes,shadows,positioning,automata,patterns}
\usetikzlibrary{trees,decorations.pathmorphing,decorations.markings}

\newcommand{\dist}{\mbox{dist}}

\newcommand{\W}{\mathcal{W}}

\newcommand{\R}{\mathbb{R}}

\newcommand{\n}{\mathbf{n}}

\newcommand{\norm}[1]{\left\Vert#1\right\Vert}

\newtheorem{thm}{Theorem}
\newtheorem{lemma}[thm]{Lemma}
\newdefinition{remark}{Remark}
\newproof{proof}{Proof}
\newproof{pot}{Proof of Theorem \ref{thm2}}

\newtheorem{problem}{Problem}
\newtheorem{condition}{Condition}

\newtheorem{definition}{Definition}
\newtheorem{example}{Example}

\newcommand{\btheorem}{\begin{thm}}
\newcommand{\etheorem}{\end{thm}}
\newcommand{\bcondition}{\begin{condition}}
\newcommand{\econdition}{\end{condition}}
\newcommand{\bremark}{ \begin{remark}}
\newcommand{\eremark}{ \end{remark}}
\newcommand{\blemma}{ \begin{lemma}}
\newcommand{\elemma}{ \end{lemma} }
\newcommand{\bdefinition}{ \begin{definition}}
\newcommand{\edefinition}{ \end{definition} }
\newcommand{\bexample}{ \begin{example}\begin{rm}}
\newcommand{\eexample}{ \end{rm}\end{example} }

\newcommand{\bproof}{\begin{proof} }
\newcommand{\eproof}{ \end{proof}}

\journal{Systems and Control Letter}

\begin{document}

\begin{frontmatter}

\title{\Large\bf Decentralized Measurement Feedback Stabilization of Large-scale Systems \\
via Control Vector Lyapunov Functions\tnoteref{t1}\footnote{A preliminary version of this paper was presented at the 2011 Australian Control
Conference (\cite{Xu2011}).}}

\author[label1]{Dabo Xu\corref{cor1}}
\ead{xu.dabo@gmail.com}
\cortext[cor1]{Corresponding author}

\author[label2]{Valery Ugrinovskii}
\ead{v.ougrinovski@adfa.edu.au}
\address[label1]{School of Automation, Nanjing University of Science and Technology, Nanjing 210094, China}
\address[label2]{School of Engineering and Information Technology, UNSW Canberra
at the Australian Defence Force Academy,\\ Po Box 7916, Canberra BC 2610, Australia}

\tnotetext[t1]{This research was supported under Australian Research
  Council's Discovery Projects funding scheme (Projects DP0987369 and
  DP120102152) and in part by National Natural Science Foundation of China under grant No. 61304009.}

\begin{abstract}
This paper studies the problem of
decentralized measurement feedback stabilization of
nonlinear interconnected systems.
As a natural extension of the recent development on control vector Lyapunov
functions, the notion of output control vector Lyapunov function (OCVLF) is introduced
for investigating decentralized measurement feedback
stabilization problems. Sufficient conditions on (local) stabilizability are
discussed which are based on the proposed notion of OCVLF. It is shown that
a decentralized controller for a nonlinear interconnected system can be constructed
using these conditions under an additional vector dissipation-like
condition. To illustrate the proposed method, two examples are given.
\end{abstract}


\end{frontmatter}

\section{Introduction}\label{SectionI}%
Large-scale system modeling has been an accepted approach to the investigation
of complex dynamical systems that consist of, or can be partitioned into
a set of interconnected subsystems. One of the most common feedback control
design strategies for such systems is the decentralized control strategy
(\cite{Lunze,Michel,Siljak91}).
Considerable efforts have been made in the literature to develop manageable
analysis and control design algorithms to reduce the computation
complexity of the existing methodologies. One such effort relates to
the notion of vector Lyapunov function~(\cite{Bellman62,Matrosov}). This
notion has been extensively used in the
analysis and control design of large-scale systems, see
\cite{KJ-book,LMS,Lunze,Michel,Siljak91,Somov}. For more recent
results on vector Lyapunov functions, we refer the reader to
\cite{Haddad04,Haddad,Karafyllis,Nersesov,Valeri}. A wide range of
applications of the method of vector Lyapunov functions to real world
problems arising in the areas of aerospace
engineering, power systems, economics, immunology, can be found in
\cite{LMS,Siljak91,Somov}.

The Lyapunov function approach dominates in the system analysis
and control theory. However in general, the construction of
a suitable (scalar) Lyapunov function for general nonlinear systems is not
a trivial task, especially when the system has a complex structure. In view
of this, the vector Lyapunov function
approach is often considered as a viable alternative to the scalar Lyapunov
method in situations involving complex systems, see
\cite{Haddad,Siljak91} for instances.
As a generalization of the standard
scalar Lyapunov function methodology, the method of vector Lyapunov functions
offers potentially more flexible strategies to cope with complexity of
dynamical
systems because it imposes different, potentially less rigid
requirements on the system components, see \cite{LMS}. Specifically, Lyapunov
functions constructed for individual subsystems of a large-scale system
only need to have certain dissipation properties. In addition, a so-called
comparison system of a reduced dimension should have certain stability property which will confirm
the corresponding stability property of the original composite system by the well-known comparison principle.

A recent development in the area of vector Lyapunov functions is concerned with the
notion of a control vector Lyapunov function and the methodology of state-feedback stabilization based on this notion, see \cite{Nersesov}. The work in
\cite{Nersesov} is an extension of the control Lyapunov
function approach originating in \cite{Artstein83,Sontag83,Sontag89}, also see
\cite{Freeman,Tsinias90,Tsinias90-2}.
Compared with these results, this paper further extends the method of
control vector Lyapunov functions to investigate problems of measurement
feedback decentralized stabilization when the complete system state information
is not available.

The main contribution of this paper is summarized
as follows. A notion of OCVLF is introduced and is used to formulate
sufficient conditions for decentralized stabilization.
We further show that, when the system has certain additional vector
dissipation properties, a constructive stabilizing control solution can be
obtained. From the theoretical viewpoint, our contribution broadens the use
of the method of vector Lyapunov
functions in the decentralized control design of large-scale nonlinear
systems.

The first result of this paper relates the existence of an OCVLF for a
nonlinear system to the existence of the partition of unity
of a certain set. In general, this makes the derivation of the stabilizing
output feedback control laws difficult in
practice, because of the lack of systematic methods to carry out the
partition of unity. Therefore, unlike the state feedback case in
\cite{Nersesov}, the computational tractability of this extension is a
critical issue. This paper shows that this issue can be circumvented in a
situation where the control input for each subsystem admits a special
decomposition into a pair of separate input
channels, cf. \cite{Tsinias90}. Specifically, we show that in this case, the
decentralized control design with an OCVLF is constructive provided the
composite system has certain vector dissipation properties.

This paper is organized as follows. Section~\ref{SectionII} describes
the class of systems under consideration and presents the formulation of the
stabilization problem for a class of large-scale interconnected
nonlinear systems that admit a certain decomposition structure. In
Section~\ref{SectionIII}, sufficient conditions for  measurement
feedback based decentralized stabilization of this class of
large-scale interconnected systems are presented.
To illustrate the proposed design method, two examples are given in
Section~\ref{Examples}. Section~\ref{Conclusion} provides concluding remarks.

{\em Notation \& Definition: } $\norm{x}$ is the Euclidean norm in $\R^n$
for $x\in\R^n$. $\R^n_+$ denotes the set of vectors with all the components
being nonnegative real numbers. In particular, $\R_+$ denotes the set of
nonnegative real numbers. For a pair of vectors $x,x'\in\R^n$, $x\prec x'$ ($x \preceq x'$ respectively) means $x_i<x'_i$
($x_i\leq x'_i$  respectively) for each $i=1,\cdots,n$. That is, $x
\preceq x'$ if and only if $x'-x\in \R^n_+$.
A function $f:\R^n\mapsto\R_+$ is said to be positive definite if $f(x)>0$
for $x\neq0$ and $f(0)=0$. The class of $k$ times continuously
differentiable functions from $\R^n$ to $\R^m$ is denoted by
$C^k[\R^n,\R^m]$, and the class of Lipschitz continuous functions is
denoted by $L[\R^n,\R^m]$. Also, $C[\R^n,\R^m]$ is the class of continuous
functions from $\R^n$ to $\R^m$.
We use $\W$ to denote the class of quasimonotone nondecreasing functions
$w(z)\in C[\R^n,\R^n]$. Recall~\cite{LMS} that a function
$w:\R^n\to\R^n$ is quasimonotone nondecreasing if for each
$i=1,\cdots,n$, $w_i(z')\leq w_i(z'')$ for any two points
$z',z''\in\R^{n}$ satisfying $z_i'=z_i''$ and $z'\preceq z''$. For two
functions $f:\R^m\to \R^l$, $g:\R^n\to \R^m$,
the notation `$\circ$' denotes the function composition, i.e., $(f\circ
g)(x) = f(g(x))$, simply denoted by $f\circ g(x)$. Given a differentiable function $W:\R^n\to \R^1$ and a
function $f:\R^n\to \R^n$, the notation $L_fW(x)$ refers to the Lie
derivative $\big(L_fW\big)(x)=\sum_{i=1}^n\frac{\partial W}{\partial x_i}f_i(x)$, simply denoted by $L_fW(x)$.


\section{Problem Formulation}\label{SectionII}%
Consider a large-scale control-affine system $\mathscr{S}$ described by the
equations
\begin{equation}
\mathscr{S} : \quad \left\{\begin{split}
\dot{x}&=f(x)+g(x)u \\
y&=h(x)
\end{split}\right.
\label{VO.S}
\end{equation}
consisting of $\n$ subsystems described by
\begin{equation*}
\mathscr{S}_i : \quad \left\{\begin{split}
\dot{x}_i& =f_i(x)+g_i(x)u_i \\
y_i&=h_i(x_i), \quad i=1,\cdots,\n.
\end{split}\right.
\end{equation*}
Here, the system state is $x=[x_1^\top,\cdots,x_{\n}^\top]^\top\in\R^n$ with
$x_i\in\R^{n_i}$, the control input is
$u=[u_1^\top,\cdots,u_{\n}^\top]^\top\in\R^m$ with $u_i\in\R^{m_i}$, the
measurement output is $y=[y_1^\top,\cdots,y_{\n}^\top]^\top\in\R^l$ with
$y_i\in\R^{l_i}$. It is assumed that $f_i,g_i,h_i$ are all smooth functions
of appropriate dimensions and
\begin{align*}
f(x)&=[f_1(x)^\top,\cdots,f_{\n}(x)^\top]^\top \\
g(x)&=\mathrm{diag}(g_1(x),\cdots,g_{\n}(x)) \\
h(x)&=[h_1(x_1)^\top,\cdots,h_{\n}(x_{\n})^\top]^\top
\end{align*}
with $f(0)=0$ and $h(0)=0$.

The large-scale system $\mathscr{S}$ will be referred to as the composite
system with a decomposition $\{\mathscr{S}_i\}_{i=1}^{\n}$. 
In view of the system structure, the subsystems $\{\mathscr{S}_i\}_{i=1}^{\n}$
are interconnected through the functions $f_i(x)$. To highlight that
the initial conditions for the system (\ref{VO.S}) are within the closed
ball $S_{\rho}=\{x\in\R^n:\norm{x}\leq\rho\}$, with a fixed real number $\rho>0$, we
will use the notation $\mathscr{S}(S_{\rho})$ for the composite system
(\ref{VO.S}).

\bremark%
It is worth noting that dynamics of each subsystem $\mathscr{S}_i$ are coupled
to other subsystems through functions $f_i,g_i$ being dependent on states
external to $\mathscr{S}_i$. On the other hand, its output $h_i$ does not
depend on external dynamics $x_j$, $j\neq i$. This model is consistent with our
objective in this paper which is to develop a methodology for
\emph{decentralized} stabilization, where the feedback law
for each subsystem $\mathscr{S}_i$ is based on its local outputs
reflecting dynamics $x_i$ of this subsystem. Problems where subsystem
measurements depend on external states $x_j$, $j\neq i$, are
usually regarded as distributed control problems. This is for example
the case in multi-agent cooperative control problems, where locally
available measurements reflect a relative state of the subsystem with
respect to its neighbors.
\eremark%

The stabilization problem for the system $\mathscr{S}( S_{\rho})$ considered
in this paper is defined as follows.

\begin{problem}[Decentralized stabilization]\label{prob-st}~%
For the given composite system $\mathscr{S}( S_{\rho})$ with the
decomposition $\{\mathscr{S}_i\}_{i=1}^{N}$, we aim to find a decentralized
controller
\begin{equation}\label{decc-1}
u=\begin{bmatrix}
u_1 \\ \vdots \\ u_{\n}  \\
\end{bmatrix}=\begin{bmatrix}
\Gamma_1\circ h_1(x_1) \\ \vdots \\ \Gamma_{\n}\circ h_{\n}(x_{\n}))  \\
\end{bmatrix}
\end{equation}
that asymptotically stabilizes the composite system $\mathscr{S}$ at the
origin. The structure of
the desired decentralized controller is illustrated
in Figure \ref{fig1}.
\end{problem}%


\tikzstyle{block}=[draw,rectangle,minimum height=1.5em, minimum width=3em]%
\begin{figure}%
 \centering%
 \begin{tikzpicture}[thick, auto, node distance=2.6cm,>=latex',scale=1.1]
 \path(4,1.7)node(S1){{\scriptsize$\mathscr{S}_1$}}%
 (9.1,1.7)node(SN) {{\scriptsize$\mathscr{S}_{\n}$}}%
 (6.5,0)node[block](p){\begin{minipage}{3.9cm}\vspace{-0.3cm}
 \begin{equation*}{\footnotesize\begin{split}
 \mathscr{S}_i:~\dot{x}_i&=f_i(x)+g_i(x)u_i \\ y_i&=h_i(x_i)
 \end{split}}\end{equation*}
 \end{minipage}}%
 (6.6,-1.511)node[block](controller){
 \begin{minipage}{.7cm}{\footnotesize$~\Gamma_i(\cdot)$}\end{minipage}};%
 \draw[->](p) -|  (8.7,0) -| (9.5,-1.511) -- node[name=yi,above left] {{\footnotesize$y_i$}}(controller);%
 \draw[->](controller) -- (3.4,-1.511) -- (3.4,0)node[name=ui,above right] {{\footnotesize$u_i$}} -- (p);%
 \draw[->](S1) edge [bend left] (4.9,0.551);%
 \draw[](5.4,1)node[name=ui,above] {$\cdots$};%
 \draw[->](5.7,1.5)node[name=ui,above] {{\scriptsize$\mathscr{S}_{i-1}$}}--(5.7,0.551)node[name=ui,above left]{{\tiny$x_{i-1}$}};%
 \draw[->](6.5,0.551)--(6.5,2.3)node[name=ui,below left]{{\tiny$x_{i}$}};%
 \draw[->](7.3,1.5)node[name=ui,above] {{\scriptsize$\mathscr{S}_{i+1}$}}--(7.3,0.551)node[name=ui,above left]{{\tiny$x_{i+1}$}};%
 \draw[](7.7,1)node[name=ui,above] {$\cdots$};%
 \draw[->](SN)edge[bend right](8.1,0.551);
 \draw[](4.6,0.8)node{{\tiny$x_1$}};%
 \draw[](8.4,0.8)node{{\tiny$x_{\n}$}};%
 \end{tikzpicture}%
\caption{Interconnection and control of subsystems}\label{fig1}%
\end{figure}
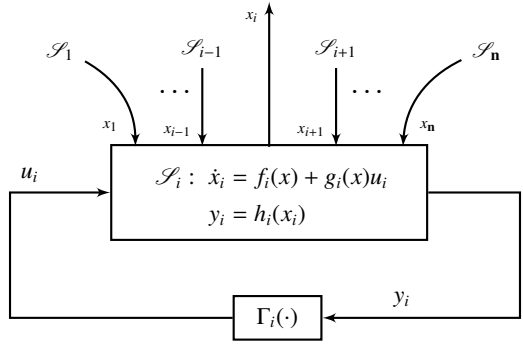%

\section{Measurement Feedback Decentralized
  Stabilization}\label{SectionIII}%
\subsection{Preliminaries}
We begin with presenting the notion of \emph{decentralized output feedback
  stabilizability} and the associated notion of OCVLF to be used in
this paper.

\bdefinition\label{defn-voutput feedback stabilizable}~%
The system $\mathscr{S}( S_{\rho})$ with a decomposition
$\{\mathscr{S}_i\}_{i=1}^{\n}$ is said to be \emph{decentralized output feedback stabilizable} (w.r.t. $ S_{\rho}$) if there exists a decentralized
controller of the form (\ref{decc-1}) that solves Problem \ref{prob-st}.
\edefinition%

Before giving the definition of an OCVLF, we introduce the following
notation. For each $i=1,\cdots,\n$ and a given set $Q\subset
\mathbb{R}^n$, let
\begin{align*}
K_i(Q)&=\big\{y_i\in\R^{l_i}: y_i=h_i(x_i),~x\in Q \big\} \\
\widetilde{K}_i(y_i)&=\big\{x\in S_{\rho}: h_i(x_i)=y_i \big\}.
\end{align*}
Note that since each $h_i$ is smooth, then $K_i(S_\rho)$ is a compact set and for each fixed $y_i\in K_i(S_\rho)$, $\widetilde{K}_i(y_i)$ is also a compact set.
Obviously, $K_i(\widetilde{K}_i(y_i))=\{y_i\}$.


\bdefinition\label{defn-OCVLF}%
The system $\mathscr{S}( S_{\rho})$ with a decomposition
$\{\mathscr{S}_i\}_{i=1}^{\n}$ is said to have an OCVLF triple
$\{V,\Lambda, S_{\rho}\}$,
where $V\in C^1[ S_{\rho},\R^{\n}_+]$ and $\Lambda\in L[\R^{\n}_+,\R^{\n}_+]\cap\W$ with
$\Lambda(0)=0$, if for each $i=1,\cdots,\n$,
\begin{itemize}
  \item [(i)] $V_i:\R^{n_i}\mapsto\R_+$ is positive definite.
  \item [(ii)] For each $y_i\in K_i( S_{\rho})\backslash\{0\}$,
there is a vector $u_i\in\R^{m_i}$ such that
\begin{equation*}
L_{f_i}V_i(x)+L_{g_i}V_i(x)\cdot u_i< \Lambda_i\circ V(x),~~ \forall x\in \widetilde{K}_i(y_i).
\end{equation*}
  \item [(iii)] For every $x\in \widetilde{K}_i(0)\backslash\{0\}$,
$L_{f_i}V_i(x)< \Lambda_i\circ V(x)$.
  \item [(iv)] The trivial solution of the following system
\begin{align}\label{z-sys}
\dot{z}&=\Lambda(z),\quad z(0)\in\R^{\n}_+
\end{align}
is asymptotically stable (cf. Definition 1.6.1 in \cite{LMS}).
\end{itemize}
\edefinition%

\begin{remark}
The properties in conditions (ii) and
    (iii) are analogous to the
    corresponding properties of scalar Lyapunov functions, and are
    fundamental for stabilizability
    by output-feedback control. The significance of these
    conditions is that they formulate the output-feedback stabilizability
    property of the system in terms of properties of individual
    subsystems. Such a formulation has proved useful in the linear case and
    the case of nonlinear systems of Lur'e type where constructive
    conditions to verify these properties were
    found~\cite{Valeri,LaU1}. Later in the
    paper, we present a constructive condition of vector dissipativity
    which addresses this question in part in the nonlinear setting.
\end{remark}

\bremark%
In the above definition, the system (\ref{z-sys}) is known as the comparison
system for $\mathscr{S}( S_{\rho})$. The key idea of the method of vector
Lyapunov functions is to reduce the stability analysis to find a
comparison system whose stability implies that of the original system.
Recently, a number of results have been developed in the literature to
facilitate the analysis of comparison systems of the form (\ref{z-sys})
associated with large-scale interconnected systems. In particular,
small-gain criteria have been developed that serve this purpose; e.g., see
\cite{Dashkovskiy07,KJ-book,Ruffer}. In the light of these results, in this
paper the comparison system will be assumed to be given.

Comparing the form of the comparison system (\ref{z-sys}) with that in
\cite{LMS,Nersesov}, we note that comparison systems could be chosen
to have a more general form, e.g., to be time-varying or
trajectory-dependent. However in this more general case, the stability of
the comparison system would need to be carefully addressed in certain
uniform sense. From this viewpoint, the results in this paper could be
further extended, at least on a case by case basis. For simplicity, we will
restrict attention to the class of time-invariant comparison systems
(\ref{z-sys}), and will use a corresponding comparison principle to be
given later in Lemma~\ref{comparison.principle}.
\eremark%

The proofs of our main results are analogous to the proofs of the similar
results in
\cite{Tsinias90,Tsinias90-2}, where
the problem of centralized measurement feedback control was studied using
a scalar Lyapunov function
in the case where $S_{\rho}$ is a small set.
For the sake of completeness, we give the full proofs to show all the
extensions.
The following lemmas will be used in the derivation of the results of
the paper. The first lemma establishes a comparison principle used in this
paper; see also Theorem 1.6.1 in \cite{LMS} or Theorem 1 in \cite{Haddad04}.
The second lemma is concerned with the
existence of a partition of unity; see page 52 in \cite{Guillemin}.
The last lemma shows that property (ii) of the OCVLF defined in
  Definition~\ref{defn-OCVLF} holds in a small neighbourhood of the point
  $y_i^{\star}$, using the same control $u_i^{\star}$. This property can be
  regarded as certain `robustness' of control $u_i^{\star}$ under small
  perturbations of $y_i^{\star}$. This robustness property will allow us to
  select a countable set of control actions $u_i^{\star}$, from which a
  smooth in $y_i$
  control law, except possibly at the origin, will be constructed using
  Lemma~\ref{PU}.

\blemma\label{comparison.principle}~%
Consider a nonlinear autonomous system described by
\begin{equation}\label{sys-Fx}
\dot{x}=\mathcal{F}(x),~~x(0)=x_0
\end{equation}
where the state is $x\in\R^n$, $\mathcal{F}\in C^1[\R^n,\R^n]$, $\mathcal{F}(0)=0$ and $x_0$ is the initial value. Suppose that
\begin{itemize}
  \item [(i)] $V\in C^1[S_{\rho},\R^{\n}_+]$ and $v(x)=\sum_{i=1}^{\n} V_i(x)$ is positive definite;

  \item [(ii)] For all $x\in S_{\rho}$, $\dot{V}(x)\preceq \Lambda\circ V(x)$,
where $\Lambda\in L[\R^{\n}_+,\R^{\n}_+]\cap\W$ and $\Lambda(0)=0$.
\end{itemize}
Also, consider the comparison system defined by
\begin{equation}\label{sys-Fx-z}
\dot{z}=\Lambda(z),~~ z(0)\in\R^{\n}_+.
\end{equation}
Then the asymptotic stability property of $z=0$ of the comparison
system (\ref{sys-Fx-z}) implies the asymptotic stability of
$x=0$ of the system (\ref{sys-Fx}).
\elemma%

Recall that for a given set $X\subset \R^n$, an open set $U_i$ is said to be a
relatively open subset with respect to $X$ if $U_i=X\cap P_i$ where
$P_i$ is some open set in $\R^n$. In this paper,  where it causes no
confusion, we will refer to relative open sets as open sets.
A collection of sets $\{U_{\alpha}\}$ covers a set $X$ if $X$ is contained in the union
$\bigcup_{\alpha} U_{\alpha}$. An open covering of $X$ is a collection of open sets $\{U_{\alpha}\}$ which covers $X$.

\blemma\label{PU}~%
Let $X$ be an arbitrary subset of $\R^n$. For any countable covering of $X$ by relatively open subsets $\{U_i\}_{i=1}^\infty$, there exists a sequence of smooth functions $\{\theta_i(x)\}_{i=1}^{\infty}$ on $X$, as a partition of unity subordinate to the open cover $\{U_i\}_{i=1}^\infty$, such that
\begin{itemize}
  \item [(i)] $0\leq\theta_i(x)\leq1$ for all $x\in X$ and all $i\geq1$.
  \item [(ii)] Each $x\in X$ has a neighborhood on which all but finitely many functions $\theta_i(x)$ are identically zero.
  \item [(iii)] Each function $\theta_i$ is identically zero except on some closed set contained in one of the $\{U_i\}_{i=1}^\infty$.
  \item [(iv)] For each $x\in X$, $\sum_{i}\theta_i(x)=1$.
\end{itemize}
\elemma%

\blemma\label{rk-neighborhood}%
For each pair $(y_i^\star,u_i^\star)$ with $y_i^\star\neq0$ satisfying
condition (ii) in Definition \ref{defn-OCVLF}, there exists an
open ball $B_{y_i^\star}$ centered
at $y_i^\star$ such that
\begin{equation}\label{b-yi0}
L_{f_i}V_i(x)+L_{g_i}V_i(x)\cdot u_i^\star< \Lambda_i\circ V(x),~~ \forall x\in \widetilde{K}_i(B_{y_i^\star})
\end{equation}
where $\widetilde{K}_i(B_{y_i^\star})\triangleq\big\{x\in S_{\rho}:h_i(x_i)\in B_{y_i^\star}\big\}$.
\elemma%

\bproof%
Fix $y_i^\star\neq0$ and $(y_i^\star,u_i^\star)$ satisfying
condition (ii) in Definition \ref{defn-OCVLF}, and introduce the set
\begin{equation*}
\widehat{K}_i^{\varepsilon}(y_i^\star)=\big\{x\in S_{\rho}: \text{dist}\big(x,\widetilde{K}_i(y_i^\star)\big)\leq \varepsilon \big\}
\end{equation*}
where $\varepsilon>0$ is a constant and
$\dist(x,\widetilde{K}_i(y_i^\star))\triangleq
\min_{s\in\widetilde{K}_i(y_i^\star)}\norm{x-s}$. Since $y_i^\star\neq0$,
due to continuity of $h_i(x_i)$ one can select a sufficiently small
$\varepsilon_1>0$ to ensure that
\begin{equation}\label{prf-epsilon00}
0\notin K_i(\widehat{K}_i^{\varepsilon}(y_i^\star)) \quad \forall
\varepsilon<\varepsilon_1.
\end{equation}

Observe that for any sufficiently small $\varepsilon>0$,
$\widehat{K}_i^\varepsilon (y_i^\star)\cap S_\rho$ is a compact
set. To establish this it suffices to show that $\widehat{K}_i^\varepsilon
(y_i^\star) \cap S_\rho$ is a closed set. Consider a converging sequence $x_k\in
\widehat{K}_i^\varepsilon (y_i^\star)\cap
S_\rho$, $\lim_{k\to \infty}x_k=x$. Since  $x_k\in
\widehat{K}_i^\varepsilon (y_i^\star)\cap
S_\rho$, then there exists $\bar x_k\in \widetilde{K}_i(y_i^\star)$ such
that $\|x_k-\bar x_k\|\le \varepsilon$. Also, since
$\widetilde{K}_i(y_i^\star)$ is compact, a converging subsequence
$\{\bar x_{k_l}\}$ can be extracted from $\{\bar x_k\}$. Let $\bar x$ be
the limit point of $\{\bar x_{k_l}\}$. We have
\begin{align*}
\|x-\bar x\|&\le
\|x-x_{k_l}\|+\|x_{k_l}-\bar x_{k_l}\|+\|\bar x_{k_l}-\bar x\| \\
&\le \varepsilon +
\|x-x_{k_l}\|+\|\bar x_{k_l}-\bar x\|.
\end{align*}
Letting $k_l\to \infty$ leads us to conclude that $\|x-\bar x\|\le
\varepsilon $. Since $\bar x\in  \widetilde{K}_i(y_i^\star)$, this implies
$x\in \widehat{K}_i^\varepsilon (y_i^\star)$. Also, $x\in S_\rho$ since the
latter set is a compact. Thus, $x\in  \widehat{K}_i^\varepsilon
(y_i^\star)\cap S_\rho$,
which confirms that  $\widehat{K}_i^\varepsilon (y_i^\star)\cap S_\rho$ is
closed. Hence, it is compact, because  $\widehat{K}_i^\varepsilon
(y_i^\star)\cap S_\rho\subseteq S_\rho $, and the latter set is bounded.

From now, let us fix $\varepsilon\in (0,\epsilon_1)$ such that
$\widehat{K}_i^\varepsilon (y_i^\star)$ is compact.

By assumption, the function $\alpha_i(x)\triangleq L_{f_i}V_i(x)+L_{g_i}V_i(x)\cdot u_i^\star-
\Lambda_i\circ V(x)$ is continuous on the compact set $S_\rho$, therefore
it is continuous on its compact subsets
$\widetilde{K}_i (y_i^\star)$ and $\widehat{K}_i^\varepsilon
(y_i^\star)$. This observation leads to the following conclusions. Firstly,
$\alpha_i(x)$ attains its maximum value on $\widetilde{K}_i (y_i^\star)$. This
implies that there exists $\delta_{y_i^\star} >0$ such that
\[
\alpha_i(x) < -\delta_{y_i^\star} \quad
\forall x\in \widetilde{K}_i(y_i^\star).
\]
Secondly, $\alpha_i(x)$ is uniformly continuous
on $\widehat{K}_i^\varepsilon (y_i^\star)$, by the Heine-Cantor Theorem. This
allows us to ascertain the existence of a sufficiently small
$\varepsilon(\delta_{y_i^\star})>0$ such that $|x-\bar x|\le
\varepsilon(\delta_{y_i^\star})$, $x,\bar x\in \widehat{K}_i^\varepsilon
(y_i^\star)\cap S_\rho$, implies
\begin{equation*}
|\alpha_i(x)-\alpha_i(\bar{x})| \le \frac{\delta_{y_i^\star}}{2}.
\end{equation*}
Since $\delta_{y_i^\star}$ and hence $\varepsilon(\delta_{y_i^\star})>0$
can be chosen to be arbitrarily small, one can always ensure that
$\varepsilon(\delta_{y_i^\star})<\varepsilon$. Therefore, if we select an
arbitrary $\bar x\in \widetilde{K}_i(y_i^\star)$, then for any $x$ such
that $\|x-\bar x\|\le \varepsilon(\delta_{y_i^\star})<\varepsilon$, it
follows that
\begin{equation}\label{VO.negative}
\alpha_i(x) \le -\frac{\delta_{y_i^\star}}{2}<0.
\end{equation}
That is, inequality (\ref{VO.negative}) holds for any
$x\in \widehat{K}_i^{\varepsilon(\delta_{y_i^\star})} (y_i^\star)$.

To complete the proof, we now show that an open ball
$B_i(y_i^\star)$ can be selected with the property that $h_i(x_i)\in
B_i(y_i^\star)$ implies
$x\in \widehat{K}_i^{\varepsilon(\delta_{y_i^\star})} (y_i^\star)$. We
prove this by contradiction. Suppose such a ball does not exist, and hence
for an arbitrarily small $\nu>0$ there exists $x_\nu$ such that
\begin{equation}\label{VO.BK}
\|h_i(x_\nu)-y_i^\star\|<\nu ~~\mbox{ and }~~ \|x_\nu- x\|>
\varepsilon(\delta_{y_i^\star}),~\forall
x\in \widetilde K(y_i^\star).
\end{equation}
The second condition means that $x_\nu\not\in \widehat{K}_i^{\varepsilon(\delta_{y_i^\star})} (y_i^\star)$.

Let us consider a sequence of radii $\nu_l=\frac{1}{l}\to 0$ as $l\to
\infty$, and let $x_{\nu_l}$ be the corresponding sequence of points
satisfying (\ref{VO.BK}). Since $\{x_{\nu_l}:l=1,2,\cdots\}\subset S_\rho$
and the latter set is compact, then a converging subsequence can be
extracted from $\{x_{\nu_l}:l=1,2,\cdots\}$, which we again denote
$\{x_{\nu_l}:l=1,2,\cdots\}$ and $\lim_{l\to\infty} x_{\nu_l}=x^0$. Due to
continuity of $h_i$, we then have
$\lim_{l\to\infty}h_i(x_{\nu_li})=h_i(x^0_i)$, and also
$\|h_i(x_{\nu_li})-y_i^\star\|<\frac{1}{l}\to 0$. Thus, $h_i(x^0_i)=y_i^\star$
due to uniqueness of the limit point. This implies that
$x^0\in  \widetilde K(y_i^\star)$. Therefore, it must follow from the second
condition (\ref{VO.BK}) that $\|x_{\nu_l}- x^0\|>
\varepsilon(\delta_{y_i^\star})$ for all $l$. However, we have previously
established that $\{x_{\nu_l}:l=1,2,\cdots\}$ converges to $x^0$. This
contradiction shows that there exists a ball $B_{y_i^\star}$ centered at
${y_i^\star}$, of sufficiently small radius, with the property
\begin{equation*}
\widetilde K_i(B_{y_i^\star})=\{x\in\R^n: h_i(x_i)\in
B_{y_i^\star}\}\subseteq\widehat{K}_i^{\varepsilon(\delta_{y_i^\star})}
(y_i^\star).
\end{equation*}
Thus, we conclude that (\ref{b-yi0}) holds.
\eproof%

\subsection{Decentralized Stabilization using an OCVLF}%

\btheorem\label{thm-1}~%
The system $\mathscr{S}( S_{\rho})$ with the decomposition
$\{\mathscr{S}_i\}_{i=1}^{\n}$ is decentralized output feedback stabilizable
by a smooth (except possibly at the origin) output feedback controller
$u=\Gamma(y)$, if there exists an OCVLF triple $\{V,\Lambda, S_{\rho}\}$
for this system.
\etheorem%

\bproof%
Suppose there is an OCVLF triple $\{V,\Lambda, S_{\rho}\}$ for
$\mathscr{S}( S_{\rho})$. Then for each $i=1,\cdots,\n$, consider the
component $V_i(x_i)$ of $V(x)$.
From condition (ii) of Definition \ref{defn-OCVLF} and by Lemma \ref{rk-neighborhood},
it follows that for
each $y_i\in K_i( S_{\rho})\backslash\{0\}$, there exist a
vector $u_i$ and an open ball $B_{y_i}$ centered
at $y_i$ such that (\ref{b-yi0}) holds.


We now note that the set $K_i(S_\rho)\backslash\{0\}$ endowed with the
Euclidean metric is a metric space. Also, this metric space is
separable. Indeed, the set $K_i(S_\rho)$ is compact, since $S_\rho$ is
compact and $h_i$ is continuous. Therefore $K_i(S_\rho)$ is separable, i.e,
it contains a dense subset. Removing, if necessary, the zero element
from this dense subset yields a dense subset for $K_i(S_\rho)\backslash\{0\}$.

Furthermore, the collection of balls $\{B_{y_i}:
y_i \in K_i(S_\rho)\backslash \{0\}$\}
forms an open covering for the separable metric space
$K_i(S_\rho)\backslash\{0\}$. According to Theorem 2 (Lindel\"{o}f) on page 94 of
\cite{Berge-1997}, it is possible to extract a
countable covering $\{B_{ij}\}_{j=1}^{\infty}$ from the covering
$\{B_{y_i^\star}: y_i^\star \in K_i(S_\rho)\backslash \{0\}\}$ such that
\begin{equation}\label{cover-condition}
K_i( S_{\rho})\backslash\{0\}\subset \bigcup_{j=1}^{\infty} B_{ij}.
\end{equation}
Here, $B_{ij}\triangleq B_{y_{ij}^\star}$ is an open ball centered at
$y_{ij}^\star\in K_i(S_{\rho})\backslash\{0\}$ satisfying (\ref{b-yi0}).
Clearly, the condition (\ref{cover-condition}) remains true when each open
set $B_{ij}$ is replaced with its relative  open version $B_{ij}\cap (K_i(
S_{\rho})\backslash\{0\})$, which we also denote $B_{ij}$.

We now observe that the conditions of Lemma~\ref{PU} are satisfied for the
set $X= K_i(S_{\rho})\backslash\{0\}$ and its relatively open covering
$\{B_{ij}\}_{j=1}^{\infty}$.  According to Lemma
\ref{PU}, there exists a sequence of smooth functions
$\{\psi_{ij}(y_i)\}_{j=1}^\infty$ with properties (i) to (iv) stated in
that lemma. In particular, we note that $\psi_{ij}(y_i)=0$ for
$y_i\notin B_{ij}$, according to claim (iii) of that lemma.

Next, define the mapping $\Gamma:\R^l\mapsto\R^m$ with $\Gamma_i:\R^{l_i}\mapsto\R^{m_i}$ described by
\begin{equation}\label{defn-Phi}
\Gamma_i(y_i)=\left\{\begin{array}{ll}
0, & \mbox{for}~y_i=0; \\
\displaystyle \sum_{j=1}^{\infty}u_{ij}^\star\psi_{ij}(y_i), &
\mbox{otherwise}
\end{array}
\right.
\end{equation}
where $u_{ij}^\star$ are vectors corresponding to the centers $y_{ij}^\star$
of the covering sets $B_{ij}$.
By virtue of property (iv) of Lemma \ref{PU}, $\Gamma_i(y_i)$ is well
defined and smooth on $K_i( S_{\rho})\backslash\{0\}$, because according to
(ii), the sum in (\ref{defn-Phi}) contains a finite number of addends for
each $y_i\in K_i( S_{\rho})\backslash\{0\}$.

Now let us fix $x\in S_\rho$, $x\neq 0$, and consider $y_i=h_i(x_i)$. Suppose
$y_i\neq 0$, then
\begin{align*}
&L_{f_i}V_i(x)+L_{g_i}V_i(x)\cdot \Gamma_i(y_i) \\
&=\Big(\sum_{j=1}^\infty
\psi_{ij}(y_i)\Big)L_{f_i}V_i(x)+L_{g_i}V_i(x)\cdot \sum_{j=1}^\infty u_{ij}^\star\psi_{ij}(y_i) \\
&=\sum_{j=1}^\infty L_{f_i}V_i(x)\psi_{ij}(y_i)+\sum_{j=1}^\infty
L_{g_i}V_i(x)\cdot u_{ij}^\star\psi_{ij}(y_i) \\
&=\sum_{j:~y_i\in B_{ij}} \Big(L_{f_i}V_i(x)+L_{g_i}V_i(x)\cdot u_{ij}^\star\Big)\psi_{ij}(y_i) \\
&\quad +\sum_{j:~y_i\not\in B_{ij}} \Big(L_{f_i}V_i(x)+L_{g_i}V_i(x)\cdot u_{ij}^\star\Big)\psi_{ij}(y_i) \\
&=\sum_{j:~y_i\in B_{ij}} \Big(L_{f_i}V_i(x)+L_{g_i}V_i(x)\cdot u_{ij}^\star\Big)\psi_{ij}(y_i).
\end{align*}
Here we used claim (iii) of Lemma~\ref{PU}, stating that $\psi_{ij}(y_i)=0$
for $y_i\not\in B_{ij}$.

Next we observe that by definition, the inclusion
$y_i=h_i(x_i)\in B_{ij}$ implies $x\in \widetilde{K_i}(B_{ij})$. Therefore $x\in
\cap_{j:~y_i\in B_{ij}}\widetilde{K_i}(B_{ij})$. Also, according to
Lemma~\ref{rk-neighborhood}, for all $j$ such that $y_i\in B_{ij}$
\[
L_{f_i}V_i(x)+L_{g_i}V_i(x)\cdot u_{ij}^\star< \Lambda_i\circ V(x) ~~\mbox{since }~~x\in
\widetilde{K_i}(B_{ij}).
\]
This allows us to conclude that
\begin{align*}
L_{f_i}V_i(x)+L_{g_i}V_i(x)\cdot \Gamma_i(y_i)
&< \sum_{j:~y_i=h_i(x_i)\in B_{ij}} \Lambda_i\circ
V(x)\cdot\psi_{ij}(y_i) \\
&\le \Lambda_i\circ V(x).
\end{align*}

We have shown that for any $x\in \widetilde{K}_i\big(K_i( S_{\rho})\backslash\{0\}\big)$
\begin{equation}\label{dV-1}
L_{f_i}V_i(x)+L_{g_i}V_i(x)\cdot \Gamma_i(y_i)< \Lambda_i\circ V(x).
\end{equation}
Also, from condition (iii) of Definition \ref{defn-OCVLF} we have for any $x\in \widetilde{K}_i(0)\backslash\{0\}$
\begin{align}\label{dV-2}
L_{f_i}V_i(x)+L_{g_i}V_i(x)\cdot \Gamma_i(0)=L_{f_i}V_i(x)&< \Lambda_i\circ V(x).
\end{align}

From (\ref{dV-1}) and (\ref{dV-2}), along the trajectory of
$\dot{x}_i=f_i(x)+g_i(x)\Gamma_i(y_i)$, $V_i(x_i)$ satisfies
\begin{equation}\label{dV-3}
\dot{V}_i(x_i)<\Lambda_i\circ V(x), \quad \forall x\in
S_{\rho}\backslash\{0\}.
\end{equation}
Therefore, together with the trivial case $x=0$, $V(x)$ satisfies the inequality
$\dot{V}(x)\preceq \Lambda\circ V(x)$ for all $x\in  S_{\rho}$.

By condition (iv) of Definition \ref{defn-OCVLF} and using Lemma
\ref{comparison.principle}, we conclude that the origin of the closed-loop
system is asymptotically stable.
\eproof%

To further characterize continuity properties of the decentralized feedback
controller of
Theorem \ref{thm-1}, we give the following definition.
\bdefinition \label{defn-vscp}~%
The system $\mathscr{S}( S_{\rho})$ with the decomposition
$\{\mathscr{S}_i\}_{i=1}^{\n}$ satisfies the (decentralized) \emph{small control property} if
for each $i=1,\cdots,\n$, there exists a continuous positive definite
function $\mu_i(y_i)\in\R_+$, with the following property:
For each $y_i\in K_i( S_{\rho})\backslash\{0\}$, there exists some
$u_i\in\R^{m_i}$ such that $\norm{u_i}<\mu_i(y_i)$ and
\begin{equation}\label{scp-eq}
L_{f_i}V_i(x)+L_{g_i}V_i(x)\cdot u_i< \Lambda_i\circ V(x),\quad\forall x\in \widetilde{K}_i(y_i).
\end{equation}
\edefinition%

In (\ref{scp-eq}) and elsewhere, we acknowledge that due to the fact that
$y_i\neq0$ in Definition \ref{defn-vscp}, $x=0$ is not contained in the set
$\widetilde{K}_i(y_i)$.

We are now in a position to present a sufficient condition for the
existence of a decentralized output feedback stabilizing controller
expressed as a continuous function.
\btheorem\label{thm-3.2-scp}~%
The system $\mathscr{S}( S_{\rho})$ is decentralized output feedback
stabilizable by a continuous decentralized output feedback control law
$\Gamma(y)$,
if there is an OCVLF triple $\{V,\Lambda, S_{\rho}\}$ for this system and moreover the small control property in the sense of Definition \ref{defn-vscp} holds.
\etheorem%

\bproof%
First, we make the following observation. From (\ref{scp-eq}) in Definition \ref{defn-vscp}, for each $y_i \in
K_i( S_{\rho})\backslash\{0\}$ with its corresponding $u_i$, there exists an open ball $B_i=B_{y_i}$ centered at
$y_i$ and satisfying the following conditions:
\begin{enumerate}[(i)]
  \item
 $L_{f_i}V_i(x)+L_{g_i}V_i(x)\cdot u_{i}< \Lambda_i\circ V(x)$, $\forall x\in \widetilde{K}_i(B_i)\backslash\{0\}$, and
  \item
 $u_i$ satisfies $\norm{u_i}<\mu_i(y_i)$,
    $\forall y_i\in B_i$.
  \end{enumerate}

Indeed, since the function $\mu_i(\cdot)$ in Definition~\ref{defn-vscp} is
continuous, for each $y_i\in K_i( S_{\rho})\backslash\{0\}$, it
follows from the condition $\norm{u_i}<\mu_i(y_i)$ that there
exists a sufficiently small open ball $U_{i1}(y_i)$ centered at $y_i$
such that $\norm{u_i}<\mu_i(y_i)$ for all $y_i\in U_{i1}$. Also, by
condition (\ref{scp-eq}) and Lemma \ref{rk-neighborhood},
there exists an open ball
$U_{i2}$ which is also centered at $y_i$ and such that
\begin{equation*}
L_{f_i}V_i(x)+L_{g_i}V_i(x)\cdot u_i< \Lambda_i\circ V(x),~~ \forall x\in
\widetilde{K}_i(U_{2i}).
\end{equation*}


Observe that both balls are centered at $y_i$. This leads us to conclude
that by choosing the smallest ball among $U_{i1}$,
$U_{i2}$ as $B_i$, we will ensure the satisfaction of both properties (i)
and (ii) stated at the beginning of the proof.

Next, in the same manner as was done in the proof of Theorem~\ref{thm-1}, a
sequence of open balls $\{B_{ij}=B_{y_{ij}^\star}\}_{j=1}^\infty$ can be
selected which satisfy (\ref{cover-condition}) and also satisfy the
corresponding versions of conditions (i) and (ii); that is, for every $j$,
there exists $u_{ij}^\star$ such that
\begin{enumerate}[(i')]
  \item
 $L_{f_i}V_i(x)+L_{g_i}V_i(x)\cdot u_{ij}^\star< \Lambda_i\circ V(x)$, $\forall x\in \widetilde{K}_i(B_{ij})\backslash\{0\}$;
  \item
$u_{ij}^\star$ satisfies $\|u_{ij}^\star\|<\mu_i(y_{i})$,
    $\forall y_i\in B_{ij}$.
\end{enumerate}

Using the above selected balls $\{B_{ij}\}_{j=1}^\infty$ and
Lemma~\ref{PU},  we can now
construct the controller $\Gamma(y)$ with
$\Gamma_i:\R^{l_i}\mapsto\R^{m_i}$ defined in (\ref{defn-Phi}).
It remains to show that the function $\Gamma_i(y_i)$ is continuous. The stability of the closed-loop system can be shown
by the same argument as in the proof of Theorem \ref{thm-1}.

It follows from (\ref{defn-Phi}) that for any $y_i\in K_i(S_{\rho})\backslash\{0\}$
\begin{equation}\label{defn-Phi.2}
\Gamma_i(y_i)= \sum_{j: y_i\in B_{ij}} u_{ij}^\star\psi_{ij}(y_i) +
\sum_{j: y_i\not\in B_{ij}} u_{ij}^\star\psi_{ij}(y_i).
\end{equation}
Using statement (iii) of Lemma~\ref{PU} concerning the partition of unity
subordinate to the covering $\{B_{ij}\}_{j=1}^\infty$, we conclude that the
second sum vanishes. On the other hand, for all $j$ such that $y_i\in
B_{ij}$, we have established that $\|u_{ij}^\star\|<\mu_i(y_i)$. Therefore,
\begin{equation}\label{continuity}
\norm{\Gamma_i(y_i)}\le \sum_{j: y_i\in B_{ij}}
\norm{u_{ij}^\star}\psi_{ij}(y_i) \leq \mu_i(y_i)
\end{equation}
where the second inequality follows from the fact that $\sum_{j: y_i\in
B_{ij}}\psi_{ij}(y_i)\le\sum_{j=1}^\infty\psi_{ij}(y_i) =1$.
Hence, $\Gamma_i(y_i)$ is continuous at $y_i=0$ and $\Gamma_i(0)=0$ because
$\mu_i(y_i)$ is continuous and positive definite at $y_i=0$. Furthermore,
according to Lemma~\ref{PU} the functions $\psi_{ij}$ are smooth. This
implies  that $\Gamma_i(y_i)$ is continuous in $K_i( S_{\rho})$ and so is
$\Gamma(y)$. The proof is complete.
\eproof%

\subsection{Constructive Decentralized Stabilization}%

As pointed out in the introduction, the controller synthesis based on
partitioning the unity is not constructive due to the lack of regular
efficient methods to compute a suitable precise partition.
This comment serves as a
motivation for the material in this section, which is focused on another
synthesis procedure. Here, we specialize the result
of Theorem \ref{thm-1} to a certain class of large-scale systems, which
admits a decomposition of the control input into a pair of separate
input channels as shown in Figure \ref{fig2}.

The introduced condition below is closely related to the vector
dissipativity theory, see \cite{Haddad04}.
The main advantage of the approach undertaken in this
section is that the decentralized design proposed here does not rely on the
partitioning the unity. It is constructive if a vector dissipation-like
condition is satisfied in addition to an existing OCVLF. Specifically, we will assume
a candidate control law for the first channel that assists the construction of the second one.

The mathematical class of systems amenable to this design method is stated
in the following condition which formulates the required vector
dissipation-like property of the system.

\tikzstyle{block}=[draw,rectangle,minimum height=1.5em, minimum width=3em]%
 \begin{figure}%
\centering%
\begin{tikzpicture}[thick, auto, node distance=2.6cm,>=latex',scale=1.1]
\path(4,1.7)node(S1){{\scriptsize$\mathscr{S}_1$}}%
(9.1,1.7)node(SN) {{\scriptsize$\mathscr{S}_{\n}$}}%
(6.5,0)node[block](p){\begin{minipage}{3.9cm}\vspace{-0.3cm}
\begin{equation*}{\footnotesize\begin{split}
\mathscr{S}_i:~\dot{x}_i&=f_i(x)+g_i(x)u_i \\ y_i&=h_i(x_i)
\end{split}}\end{equation*}
\end{minipage}}%
(6.5,-1.1)node[block](controller1){
\begin{minipage}{.7cm}{\footnotesize$\phi_{i1}(\cdot)$}\end{minipage}}%
(6.5,-1.8)node[block](controller2){
\begin{minipage}{.7cm}{\footnotesize$\phi_{i2}(\cdot)$}\end{minipage}};%

\draw[->](9.5,-1.1) -- node[name=yi,above left] {{\footnotesize$y_i$}}(controller1);%
\draw[->](controller1) -- (3.7,-1.1) -- (3.7,-.2)node[name=ui,below right] {{\footnotesize$u_{i1}$}}-- (4.63,-.2);%
\draw[->](p) -|  (8.7,0) -| (9.5,-1.8) -- node[name=yi,above left] {{\footnotesize$y_i$}}(controller2);%
\draw[->](controller2) --(3.4,-1.8) -- (3.4,.2)node[name=ui,above right] {{\footnotesize$u_{i2}$}} -- (4.63,.2);%
\draw[->](S1) edge [bend left] (4.9,0.551);%
\draw[](5.4,1)node[name=ui,above] {$\cdots$};%
\draw[->](5.7,1.5)node[name=ui,above] {{\scriptsize$\mathscr{S}_{i-1}$}}--(5.7,0.551)node[name=ui,above left]{{\tiny$x_{i-1}$}};%
\draw[->](6.5,0.551)--(6.5,2.3)node[name=ui,below left]{{\tiny$x_{i}$}};%
\draw[->](7.3,1.5)node[name=ui,above] {{\scriptsize$\mathscr{S}_{i+1}$}}--(7.3,0.551)node[name=ui,above left]{{\tiny$x_{i+1}$}};%
\draw[](7.7,1)node[name=ui,above] {$\cdots$};%
\draw[->](SN)edge[bend right](8.1,0.551);
\draw[](4.6,0.8)node{{\tiny$x_1$}};%
\draw[](8.4,0.8)node{{\tiny$x_{\n}$}};%
\end{tikzpicture}%
\caption{A two-channel control system configuration.}\label{fig2}%
\end{figure}
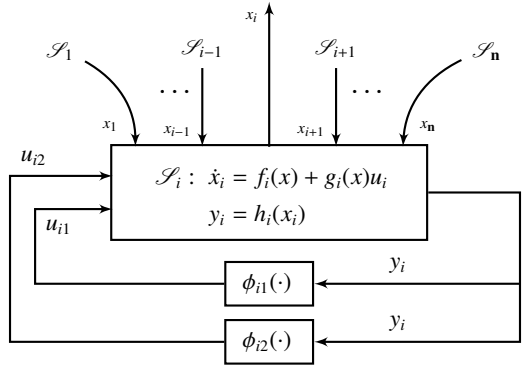%

\bcondition\label{A-construct}%
The system $\mathscr{S}( S_{\rho})$ has a decomposition
$\{\mathscr{S}_i\}_{i=1}^{N}$, and its input $u_i$ is decomposed into
$u_i=(u_{i1},u_{i2})^\top$, $u_{ij}\in\R^{m_{ij}}$, $j=1,2$ and
$m_{i1}+m_{i2}=m_i$ for $i=1,\cdots,\n$ (one of the components can be of
zero dimension). Furthermore, there exist functions
$V\in C^1[S_\rho,\R^{\n}_+]$, whose components $V_i(x)=V_i(x_i)$,
$i=1,\ldots,N$, are all positive definite,
and $\Lambda\in L[\R^{\n}_+,\R^{\n}_+]\cap \mathcal{W}$, with
$\Lambda(0)=0$,
such that
\begin{enumerate}[(i)]
  \item The trivial solution of the system $\dot z =\Lambda(z)$ is asymptotically
    stable.

  \item $V_i$ satisfies, along the trajectory of $\mathscr{S}_i$, the following dissipation-like inequality
\begin{equation}
\dot{V}_i(x_i)\leq W_i(x,u_{i1})+p_{i1}(y_i,u_{i1})+p_{i2}(y_i)u_{i2}
\label{VO.***}
\end{equation}
for some function $W_i(x,u_{i1})\in\R$ with $W_i(0,0)=0$ and smooth functions $p_{i1}(y_i,u_{i1})\in\R$ with $p_{i1}(0,0)=0$ and $p_{i2}^\top(y_i)\in\R^{m_{i2}}$.

Here and in the following, along $\mathscr{S}_i$, it means $\dot{V}_i(x_i)=L_{f_i}V_i(x)+L_{g_i}V_i(x)\cdot u_i$.

  \item There exists a smooth control law $u_{i1}=\phi_{i1}(y_i)$
    vanishing at the origin such that
\begin{equation}
W_i(x,\phi_{i1}(y_i))\leq \Lambda_i\circ V(x),~~\forall x\in  S_{\rho}.
\label{VO.**}
\end{equation}

  \item $p_{i1}(y_i,\phi_{i1}(y_i))\leq 0$
for any $y_i\in\big\{y_i\in K_i( S_{\rho}):p_{i2}(y_i)=0\}$.
\end{enumerate}
\econdition%

Condition \ref{A-construct} alludes to a two-step controller design
procedure for the system whose subsystems have a two-channel
structure shown in Figure~\ref{fig2}. First, using the function $W$ defined
in (ii), a control low $u_{i1}=\phi_{i1}(y_i)$ is obtained to satisfy condition
(\ref{VO.**}). This ensures that the composite system resulting from
closing the inner feedback loop is vector dissipative
with respect to the vector supply rate
\begin{equation*}
S(u_{i2},y_i)=p_{i1}(y_i,\phi_{i1}(y_i))+p_{i2}(y_i)u_{i2}
\end{equation*}
see Definition~6 in~\cite{Haddad04} for the definition of the notion of vector
dissipativity. This property is a natural extension of the corresponding
scalar dissipativity property, see \cite{Isidori99}. Next, at the second step,
the design of $u_{i2}$ is carried out to ensure that, for any $y_i\neq0$
\begin{equation*}
S(\phi_{i2}(y_i),y_i)=p_{i1}(y_i,\phi_{i1}(y_i))+p_{i2}(y_i)\phi_{i2}(y_i)<0.
\end{equation*}
Due to the above mentioned vector dissipativity property and condition (i),
this controller $u_{i2}=\phi_{i2}(y_i)$ will ensure that the closed-loop
system is asymptotically stable at the origin.

\bremark\label{remark5}%
Note that this procedure and Condition \ref{A-construct} can also be
adapted for a special case where the inner loop in Figure~\ref{fig2} is absent, and $u_i=u_{i2}$. In this case, conditions (ii) and (iii) can be combined into
one condition
\begin{equation*}
\dot{V}_i(x_i)\leq\Lambda_i\circ V(x)+p_{i1}(y_i)+p_{i2}(y_i)u_i
\end{equation*}
which imposes on the system the vector dissipativity requirement with the
vector supply rate
\begin{equation*}
S(u_i,y_i)=p_{i1}(y_i)+p_{i2}(y_i)u_i.
\end{equation*}
\eremark%

The following theorem crystallizes the above discussion in the form of a
concrete stabilization algorithm.

\btheorem~\label{thm-3.3}%
The composite system $\mathscr{S}( S_{\rho})$ is decentralized output feedback
stabilizable if Condition \ref{A-construct} is satisfied. Furthermore, one
stabilizing controller for this system is given by
\begin{equation}\label{thm3.3-u}
u_i=\phi_i(y_i)=\begin{bmatrix}
\phi_{i1}(y_i) \\ \phi_{i2}(y_i) \\
\end{bmatrix}.
\end{equation}
In (\ref{thm3.3-u}), $\phi_{i1}(y_i)$ is the function with properties
(iii), (iv) of Condition  \ref{A-construct},
and $\phi_{i2}(y_i)$ is defined as follows
\begin{equation}\label{thm3.3-varphi2-1}
\phi_{i2}(y_i)=\left\{\begin{array}{ll}
0, & \mbox{if}~ y_i=0; \\
\varphi_i\big(\tilde{p}_{i1},p_{i2},\sigma_i\big)\cdot p_{i2}^{\top}, & \mbox{otherwise}
\end{array}\right.
\end{equation}
where $\tilde{p}_{i1}(y_i)\triangleq p_{i1}(y_i,\phi_{i1}(y_i))$, $\sigma_i: \mathbb{R}^{l_i}\mapsto\mathbb{R}_+$ is a smooth nonnegative design function vanishing at $y_i=0$, and
\begin{align}\label{thm3.3-varphi2-i}
\varphi_i(\tilde{p}_{i1},p_{i2},\sigma_i)&=\left\{\begin{array}{ll}
0, & \mbox{if}~p_{i2}(y_i)=0; \\
-\frac{\tilde{p}_{i1}+\sigma_i}{\norm{p_{i2}}^2}, &
\mbox{otherwise}.
\end{array}\right.
\end{align}
\etheorem%

\bproof%
Consider the closed-loop system composed of $\mathscr{S}(S_{\rho})$ and (\ref{thm3.3-u}). From conditions (ii) to (iv) of Condition \ref{A-construct}, for each
$i=1,\cdots,\n$, $V_i(x)$ satisfies, along the trajectory of $\mathscr{S}_i$
with $u_i=\phi_i(y_i)$, the following inequality
\begin{align*}
\dot{V}_i(x_i)&\le W_i(x,\phi_{i1}(y_i))+p_{i1}(y_i,\phi_{i1}(y_i)) +p_{i2}(y_i)\phi_{i2}(y_i) \\
&\leq\left\{\begin{array}{ll}
\Lambda_i\circ V(x), & \mbox{if~} y_i=0 \mbox{~or~} p_{i2}(y_i)=0; \\
\Lambda_i\circ V(x)-\sigma_i(y_i), &  \mbox{otherwise} \end{array}\right. \\
&\leq \Lambda_i\circ V(x).
\end{align*}
Therefore, we have
$\dot{V}(x)\preceq \Lambda\circ V(x)$.
Then, using condition (i) of Condition~\ref{A-construct} and  the
comparison principle, we conclude that $\mathscr{S}( S_{\rho})$ is
decentralized output feedback stabilizable.
\eproof%

\bremark\label{songtag-form}%
The function $\sigma$ in Theorem~\ref{thm-3.3} is the design parameter
which provides a certain
freedom at the second stage of the design algorithm. For example, the
flexibility in selecting the design function $\sigma_i(y_i)$ allows us to adopt
the Sontag formula to obtain $u_{i2}$ (see \cite{Sontag89}). Indeed,
selecting the design function $\sigma_i(y_i)$  to be
\begin{equation}\label{thm3.3-varphi2-SA}
\sigma_i(y_i)= \sqrt{\tilde{p}_{i1}^2+\norm{p_{i2}}^2}
\end{equation}
yields the aforementioned controller $u_{i2}$ for $\mathscr{S}(
S_{\rho})$. The complete control law is then given by equations
(\ref{thm3.3-varphi2-1}) , (\ref{thm3.3-varphi2-i}), and
(\ref{thm3.3-varphi2-SA}). Alternatively, $\sigma_i(y_i)$ can be chosen
to be equal to zero.
\eremark%

\bremark%
We also note that the proposed theory can be extended to include
vector storage functions $V_i$ as components of $V$. In this case, each
inequality
in Condition \ref{A-construct} needs to be understood as a coordinate-wise
inequality. In some situations, particularly for large-scale systems
composed of structured subsystems (i.e. when the system has a nested
structure), such an extension may offer some convenience.
Therefore, our treatment utilizing a scalar storage function for
each subsystem is quite general. 
\eremark%


\section{Examples}\label{Examples}
We now present examples to illustrate the theory developed in the
preceding sections. In the first example, we illustrate the situation
where, while the OCVLF $V$ exists, output feedback stabilizability does not
automatically follow using a trivial Lyapunov function $V=\sum_{i=1}^\n
V_i(x_i)$. The second example illustrates the construction of a
decentralized controller by applying Theorem \ref{thm-3.3} with a vector
storage function for each subsystem.

\bexample

Consider the following system
\begin{equation}\label{lorenz-sim}
\left\{\begin{split}
&\dot{x}_{1}=-x_{1}+2x_1x_{3}^2 \\
&\dot{x}_{2}=x_{1}-x_{2}-x_{1}x_{3}+u_2 \\
&\dot{x}_{3}=x_{1}x_{2}-x_{3}.
\end{split}\right.
\end{equation}
In order to demonstrate that this system fits into the framework of our
theory, we will treat each equation as a subsystem and define subsystems
outputs to be  $y_i=x_{i},~i=1,2,3$. Also, we will assume that
$u_1=u_3\equiv0$.

We now demonstrate that the function
\begin{equation}\label{lorenz-simV1}
V(x)=[V_1,V_2,V_3]^\top,\quad V_i=\frac{1}{2}x_i^2,\quad i=1,2,3
\end{equation}
is a valid OCVLF for the system
(\ref{lorenz-sim}) in the sense of Definition~\ref{defn-OCVLF}, while
\begin{equation}\label{lorenz-simV2}
\bar{V}(x)=\sum_{i=1}^3 V_{i}
\end{equation}
fails to satisfy the conditions for output-feedback stabilizability
in~\cite{Tsinias94}.

Suppose that $S_{\rho}=\{x\in\R^3:\norm{x}\leq2\}$ in this example. It can
be readily verified that in $S_{\rho}\backslash \{0\}$ the functions
(\ref{lorenz-simV1}) satisfy the following conditions
\begin{align*}
\dot{V}_1&=-x_{1}^2+2x_1^2x_{3}^2  \\
&\leq -2V_1+8V_3 \\
&<-(2-\epsilon) V_1+\epsilon V_2+ (8+\epsilon) V_3, \\
\dot{V}_2&= x_2x_{1}-x_{2}^2-x_{1}x_2x_{3}+x_2u_2\\
&\leq \frac{1}{2}x_{1}^2+\frac{1}{2}x_{2}^2-x_{2}^2+2x_2^2+\frac{1}{2}x_{3}^2+x_2u_2  \\
&=V_1+3V_2+ V_3+x_2u_2 \\
&<(1+\epsilon)V_1+(3+\epsilon)V_2+(1+\epsilon) V_3 + x_2u_2,\\
\dot{V}_3&= -x_{3}^2+x_1x_{2}x_{3} \\
&\leq 4V_2-V_3\\
&< \epsilon V_1+(4+\epsilon)V_2-(1-\epsilon)V_3.
\end{align*}
Letting $u_2=-\kappa y_2$ yields the vector inequality
\begin{equation*}
\dot{V}\prec \Lambda V, \quad \Lambda\triangleq\begin{bmatrix}
-2+\epsilon & \epsilon & 8+\epsilon \\ 1+\epsilon & 3+\epsilon-\kappa & 1+\epsilon \\ \epsilon & 4+\epsilon & -1+\epsilon
\end{bmatrix}.
\end{equation*}
Take $\kappa=33$, $\epsilon=0.001$. Then, $-\Lambda$ is an M-matrix and
$\Lambda$ is
Hurwitz. This observation verifies conditions (ii) and (iv) of
Definition~\ref{defn-OCVLF}.

Next we verify condition (iii). For $i=1$ and $y_1=x_1=0$, we have
$\widetilde{K}_1(0)\backslash\{0\}=\{x\in\R^3:x_1=0,~0<x_2^2+x_3^2\le 4\}$ and
$V_1(x_1)=0$ for all $x\in\widetilde{K}_i(0)\backslash\{0\}$. This ensures
that
\begin{equation*}
\dot V_1 =0 < \Lambda_1V, \quad \forall x\in\widetilde{K}_1(0)\backslash\{0\}
\end{equation*}
where $\Lambda_1$ denotes the first row of the matrix $\Lambda$.
The cases $i=2,3$ are considered in a similar manner. Hence condition (iii)
is satisfied, which leads us to conclude that the function
(\ref{lorenz-simV1}) is an OCVLF.

However, it can be seen that (\ref{lorenz-simV2}) satisfies, along (\ref{lorenz-sim})
\begin{align*}
\dot{\bar{V}}&
=-x_1^2+2x_1^2x_3^2-x_3^2+x_1y_2x_3+y_2(x_{1}-y_2-x_{1}x_{3})+y_2u.
\end{align*}
Clearly when $y_2=0$, we have $\dot{\bar{V}}=0$ at $x_1=x_3=1$, that is, the
function (\ref{lorenz-simV2}) fails to satisfy Condition C) in
\cite{Tsinias94}. Hence, the function (\ref{lorenz-simV2}) cannot be used
as a scalar control Lyapunov function for output-feedback stabilization
of the system (\ref{lorenz-sim}).
\eexample

\bexample%
Adopted from \cite{Duan09}, consider a network of controlled Lorenz-type
systems described by
\begin{equation}\label{Lorenz1}
\mathscr{S}_i : \quad \left\{\begin{split}
\dot{x}_{i1} &=w_1(x_{i2}-x_{i1}) \\
\dot{x}_{i2} &=w_3x_{i1}-x_{i2}-x_{i1}x_{i3}+u_i+\varpi_i\sum_{j=1,j\neq i}^\n Hx_j  \\
\dot{x}_{i3} &=x_{i1}x_{i2}-w_2x_{i3} \\
y_i&=x_{i2}, \quad i=1,\cdots,\n
\end{split}\right.
\end{equation}
where $x_i=[x_{i1},x_{i2},x_{i3}]^\top$ is the state variable of the $i$-th subsystem, $u_i$ is the control input, $H$ is a coupling matrix, and $\varpi_i>0$ is the coupling strength.
The problem here is to stabilize the system at the origin $x=0$.

\begin{figure}
  \centering%
  \includegraphics[width=7.8cm,height=7cm]{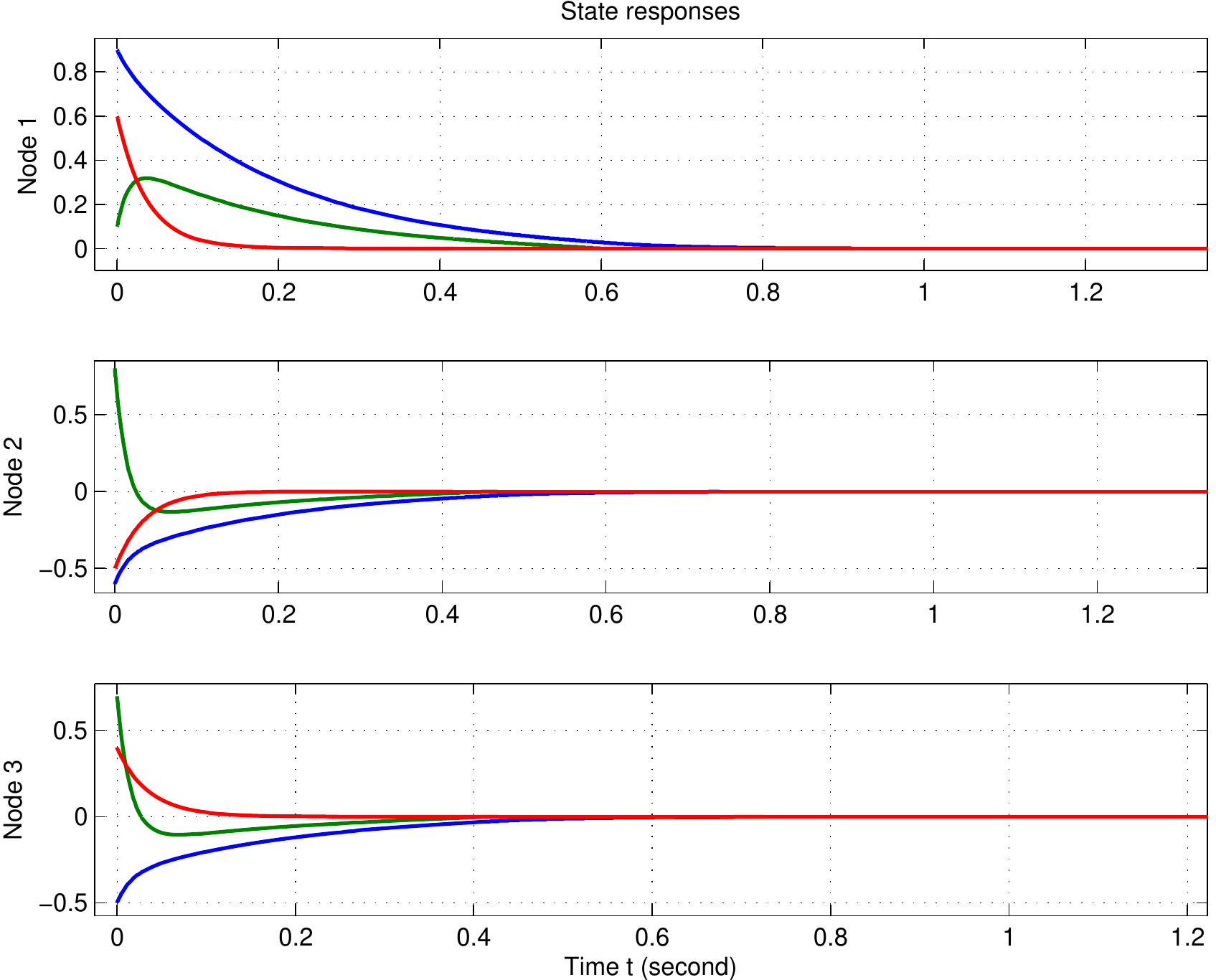}\\
  \caption{State responses of the closed-loop system with $\sigma_i$ given by the formula (\ref{thm3.3-varphi2-SA}).}\label{figeg1}
\end{figure}

In this example, we focus on a $y$-coupled type network (see Remark 1 in
\cite{Duan09}), i.e., we assume
\begin{equation*}
\varpi_i\sum_{j=1,j\neq i}^\n Hx_j=\varpi_i\sum_{j=1,j\neq i}^\n y_j.
\end{equation*}
Also, we take $w_1=10$, $w_2=\frac{8}{3}$ and $w_3=28$.
We now verify Condition \ref{A-construct}, this will ensure that Theorem
\ref{thm-3.3} is applicable to this problem.

First, for each subsystem, we construct a vector storage function. Similar to \cite{Xu2010}, notice that the $(x_{i1},x_{i3})$ subsystem
\begin{equation*}
\left\{\begin{split}
\dot{x}_{i1} &=w_1(x_{i2}-x_{i1}) \\
\dot{x}_{i3} &=x_{i1}x_{i2}-w_2x_{i3}
\end{split}\right.
\end{equation*}
admits a Lyapunov-like function of the form
\begin{equation*}
V_{i1}=\frac{1}{2}x_{i1}^2+\frac{1}{4}x_{i1}^4+\frac{1}{2}x_{i3}^2
\end{equation*}
which satisfies the dissipation inequality
\begin{equation}\label{eg1-inequ1}
\dot{V}_{i1}\leq -c_{i1}V_{i1}+c_{i2}(y_i^2+y_i^4)
\end{equation}
for some real numbers $c_{i1},c_{i2}>0$. Then, we define a vector storage
function candidate to be
\begin{equation*}
V_{i}=\begin{bmatrix}
V_{i1} \\ V_{i2} \\
\end{bmatrix}
=\begin{bmatrix}
\frac{1}{2}x_{i1}^2+\frac{1}{4}x_{i1}^4+\frac{1}{2}x_{i3}^2 \\
\frac{1}{2}y_i^2 \\
\end{bmatrix}.
\end{equation*}
Using (\ref{eg1-inequ1}), it can be established that for all $x\in S_{\rho}$
\begin{equation}
\dot{V}_{i}\preceq\begin{bmatrix}
-c_{i1}V_{i1}+c'_{i2}V_{i2} \\
y_iu_i+\frac{c_{i1}}{2}V_{i1}+c'_{i2}V_{i2}+\sum_{j=1,j\neq i}^{\n}\varpi_{ij}V_{j2}  \\
\end{bmatrix}
\label{VO.*}
\end{equation}
where $\varpi_{ij}$, $c'_{i2}$ are some appropriately chosen positive real
numbers. This condition can be written in the form of inequality
(\ref{VO.***}) in Condition~\ref{A-construct}. To show that define
$W_i=W_i(x)\in\R^2$ as follows
\begin{equation*}
W_i=\begin{bmatrix}
-c_{i1} & c'_{i2} \\
\frac{c_{i1}}{2} & -2k_i+c'_{i2}  \\
\end{bmatrix}\begin{bmatrix}
V_{i1} \\ V_{i2} \\
\end{bmatrix}+\begin{bmatrix}
0 \\
\sum_{j=1,j\neq i}^{\n}\varpi_{ij}V_{j2}  \\
\end{bmatrix}.
\end{equation*}
Notice that the control input of each subsystem in this example is
one-dimensional, that is, $u_{i1}$ is a zero-dimensional vector as
explained in Remark~\ref{remark5}. For that reason, $W_i$ is defined as a
function of $x$ only. Also, let $p_{i1}=k_iy_i^2$ and $p_{i2}=y_i$.
With these definitions, (\ref{VO.*}) can be written in
the form of inequality (\ref{VO.***}):
\begin{equation*}
\dot{V}_i(x_i)\preceq W_i(x,0)+\begin{bmatrix}
0 \\ p_{i1}(y_i,0)  \\
\end{bmatrix}+\begin{bmatrix}
0 \\ p_{i2}(y_i)u_{i}  \\
\end{bmatrix}.
\end{equation*}
This verifies conditions (ii) and (iv) of Condition \ref{A-construct}.


Next, let us consider the remaining conditions (i) and (iii) of Condition
\ref{A-construct}. It can be seen that the above defined function $W_i$
satisfies inequality (\ref{VO.**}), where $\Lambda(z)$ is a linear function
$\Lambda z$,
\begin{equation*}
\Lambda=\begin{bmatrix}
-c_{11} & c'_{12} & \cdots & 0 & 0  \\
\frac{c_{11}}{2} & -2k_1+c'_{12} & \cdots & 0 & \varpi_{1\n}  \\
\vdots & \vdots & & \vdots & \vdots  \\
0 & 0 & \cdots & -c_{\n1} & c'_{\n2} \\
0 & \varpi_{\n\n} & \cdots & \frac{c_{\n1}}{2} & -2k_\n+c'_{\n2} \\
\end{bmatrix}.
\end{equation*}
This verifies condition (iii). Also, the matrix $-\Lambda$ is an M-matrix
and $\Lambda$ can be made Hurwitz by choosing the constants $k_i$ to be sufficiently
large~\cite{Horn}. Specifically, the matrix $\Lambda$ is
Hurwitz if $k_i$ is chosen so that
\begin{equation*}
k_i>\frac{c_{i1}}{2}+c'_{i2}+\sum_{j=1,j\neq i}\varpi_{ij}.
\end{equation*}
This selection of $k_i$ verifies property (i) of Condition
\ref{A-construct}.

We have verified that the candidate OCVLF
\begin{equation}\label{eg-Vcandidate}
V=(V_1,\cdots,V_\n)^\top
\end{equation}
satisfies Condition~\ref{A-construct}. Hence, we conclude from Theorem \ref{thm-3.3} that any controller of the form
\begin{equation}\label{egc1}
u_i=\phi_{i2}(y_i)=\left\{\begin{array}{ll}
0, & \mbox{if}~ y_i=0; \\
-\frac{1}{y_i^2}\cdot (k_iy_i^2+\sigma_i)\cdot y_i, & \mbox{otherwise}
\end{array}\right.
\end{equation}
is a decentralized stabilizing controller for the system in this example.
In particular, choosing $\sigma_i=0$ yields a special controller as follows
\begin{equation}\label{egc2}
u_i=-k_iy_i
\end{equation}
which is clearly smooth in $y_i$ at the origin.

\begin{figure}
  \centering%
  \includegraphics[width=7.8cm,height=7cm]{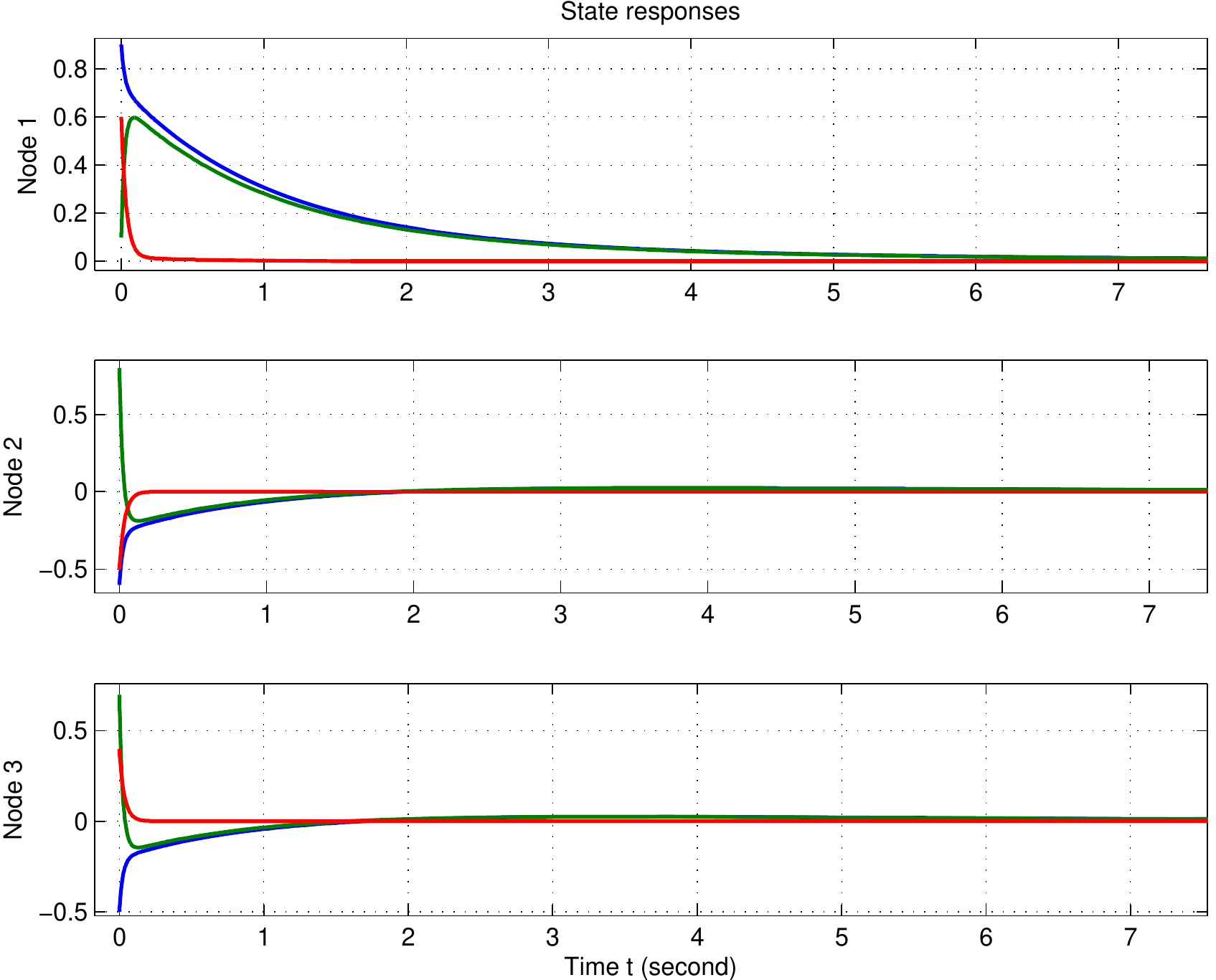}\\
  \caption{State responses of the closed-loop system with $\sigma_i=0$.}\label{figeg2}
\end{figure}

To confirm these findings, the system (\ref{Lorenz1}) was simulated with
the controller obtained using $\sigma_i$ specified in
(\ref{thm3.3-varphi2-SA}) and also with the controller (\ref{egc2}).
Simulation results for
these controllers are shown in Figures \ref{figeg1} and \ref{figeg2},
respectively. In both cases, the system parameters were selected to be
$\rho=2$, $k_i=30$, $\varpi_i=1$, $\n=3$, and the initial condition $(0.9,
0.1, 0.6; -0.6, 0.8, -0.5; -0.5, 0.7, 0.4)$ was chosen.
From these two plots, one can observe that the first controller has a
better performance, in particular, a better settling time, thanks to
selecting the design function $\sigma_i$ given by (\ref{thm3.3-varphi2-SA}).

To conclude this example, we point out that the proposed OCVLF design
procedure may potentially be used to solve some other problems in \cite{Duan09}
involving other types of coupling, as well as global stabilization or
synchronization problems. This issue is left for future research.
\eexample%

\section{Conclusion}\label{Conclusion}%
A problem of decentralized stabilization of large-scale systems via static
measurement feedback has been studied in this paper using the method of
output control vector Lyapunov functions. We have proved a general result
relating stabilizability of a large-scale system to
the existence of such a function. We also extended the results in
\cite{Nersesov,Tsinias90,Tsinias90-2} to the case of output-feedback
decentralized stabilization. A constructive design was then presented based
on a vector dissipation-like condition.
The proposed method has been applied to decentralized control of a network
consisting of a set of coupled Lorenz-type systems.

\section*{Acknowledgement}
D. Xu would like to thank his advisor Prof. Jie Huang for leading him to the field of nonlinear control and
especially the kind guidance on \cite{Artstein83,Freeman,Isidori99,Sontag89} when he began his study at The Chinese University of Hong Kong.

\end{document}